# String Models for Ultrarelativistic Hadronic Interactions: Concepts, Limitations, and New Directions [*]

K. WERNER [†]

Institut für Theoretische Physik, Universität Heidelberg
Philosophenweg 19, 69120 Heidelberg, Germany [‡]

## Abstract

After a brief description of the basic theoretical concepts of the string model approach, we discuss the limitations of the method, at low energies, at high energies, and at high particle densities. We also report on recent efforts to overcome these limitations.

## 1  Introduction

During the past decade the string model has been quite successfully applied to describe hadronic interactions, in particular nucleus-nucleus collisions at CERN-SPS energies. Awaiting experiments with larger nuclei at higher energies in the future, it appears useful to reflect about the limitations of the appraoch and about the efforts to go beyond.

In chapter 2, we present some overview of modelling hadronic interactions, also to clarify what we actually mean by "string model approach" and to show relations to other models. In chapter 3 we discuss the basic concepts of the string model, with emphasis on the way the string is introduced, namely as a parametrization of squared amplitudes and not as an elementary object. We then discuss in chapter 4 the limitations of the approach, and recent efforts to go beyond.

---

[*]Invited Lecture at the International Workshop on Multi-Particle Correlations and Nuclear Reactions, CORINNE II, 6.-10. Sept. 1994, Nantes, France
[†]Heisenberg fellow
[‡]Internet: werner@minnie.mpi-hd.mpg.de

0

## 2 Overview

When modelling ultrarelativistic hadronic interactions, one first needs to specify whether soft or hard interactions are to be considered. "Hard" refers to scatterings involving a large momentum transfer – large enough to allow for a perturbative treatment, otherwise, for small momentum transfer, the term "soft" is used. Whether a particle originates from soft or hard scattering, can be most easily seen by inspecting transverse momentum ($p_t$) spectra. Pion spectra typically show an exponential low-$p_t$ behaviour (for $p_t < 2$ GeV/c) and a large-$p_t$ behaviour of the form $p_t^{-n}$ (see fig. 1). Such a power law behaviour is typical for hard processes, described in terms of elementary QCD diagrams, whereas the exponential low-$p_t$ part is due to soft processes. The relative weight of hard processes is energy dependent (see fig. 2): at SPS energies (20 GeV) hard scattering plays only a marginal role, at LHC (6300 GeV) hard scattering is crucial. At RHIC energy (200 GeV), hard scattering is important, however, the interaction is still most likely soft. For the following, we restrict ourselves to soft interactions, keeping in mind that this provides a description up to moderately high energies only ($< 100$ GeV), but needs extension to go beyond.

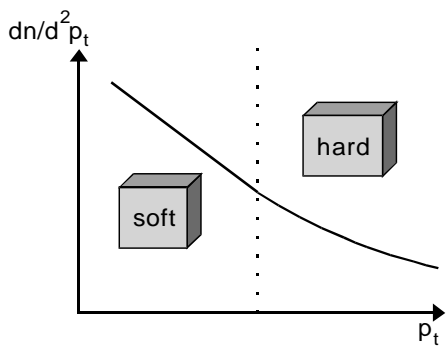
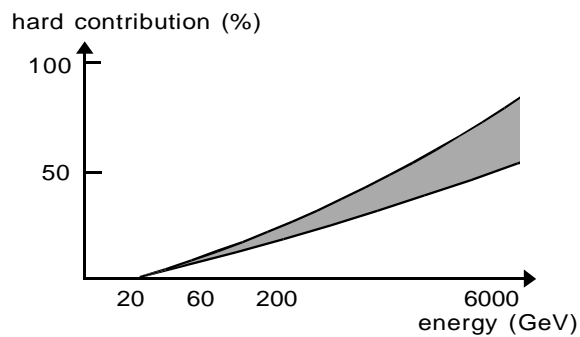

**Figure 1**: Tranverse momentum distribution separating soft and hard scattering.

**Figure 2**: Realative weight of hard scattering versus energy.

| models for UR interactions | | | |
|---|---|---|---|
| for soft interactions | | | for hard interactions |
| MD models | phenom. string models | GRT based string models | PQCD models |

**Table 1**: Classification of models for ultrarelativistic (UR) hadronic interactions

There are few different approaches to describe soft interactions, hadron-hadron as well as hadron-nucleus and nucleus-nucleus scattering (see table 1). Models based on molecular dynamics (MD) are a straightforward generalization from lower energies [1, 2]. On the other hand, there are string models, either properly defined in the frame work of relativistic quantum theory – based on Gribiv-Regge-Theory (GRT) [3, 4, 5, 6, 7] – or string models introduced in a more phenomenological fashion [8]. We do not want to comment on MD models nor on phenomenological string models, we rather provide a critical discussion of GRT-based string models, its theoretical justification and apparent limitations. So in the following, "string model" refers solely to GRT-based string models.

## 3 The String Model of Hadronic Interactions

In the following we discuss the theoretical concepts of the string model approach, implemented in models like VENUS, DPM, and QGSM. One first considers the elastic amplitude $A(s,t)$ for hadron-hadron scattering, with $s$ and $t$ being the Mandelstam variables $s = (p_1 + p_2)^2$ and $t = (p_1 - p_3)^2$, where $p_1, p_2$ are the incoming and $p_3, p_4$ the outgoing momenta. The amplitude $A(s,t)$ is given as a multiple Pomeron exchange series, with the zigzag lines symbolizing Pomerons. The Pomeron is the elementary exchange object here, without being elementary in terms of quarks and gluons. Originally the Pomeron was thought to be a gluon ladder. According to Veneziano, a Pomeron is the sum of all QCD diagrams of cylindrical topology A gluon ladder is, by the way, of cylindrical topology, in some sense the simplest cylinder.

Whatever the precise nature of the Pomeron may be: first of all one simply parametrizes the Pomeron propagator as

$$G(s,t) \sim s^{\alpha(t)} = s^{\alpha(0)+\alpha't}, \tag{1}$$

with this Regge pole form coming from general considerations of amplitudes in the limit $s \to \infty$. Two parameters characterize the Regge trajectory $\alpha(t)$: the "intercept" $\alpha(0)$ and the "slope" $\alpha'$, both being adjusted to fit data. With this simple form for $G$, the convolutions $G \otimes ... \otimes G$ can be worked out and summed over, and one obtains

$$A(s,t) = \frac{i}{4\pi} \int d^2b \, e^{i\vec{k}\vec{b}} \gamma(s,b), \tag{2}$$

with

$$\gamma(s,b) = 1 - e^{-\omega(s,b)}, \tag{3}$$

|  | 19.4 GeV | 200 GeV | 6.3 TeV |
|---|---|---|---|
| $w_1$ | 0.54 | 0.52 | 0.47 |
| $w_2$ | 0.24 | 0.23 | 0.22 |
| $w_3$ | 0.12 | 0.13 | 0.13 |
| $w_4$ | 0.059 | 0.067 | 0.081 |
| $w_5$ | 0.026 | 0.033 | 0.047 |
| $w_6$ | 0.011 | 0.015 | 0.026 |
| $w_7$ | 0.0040 | 0.0061 | 0.0130 |
| $w_8$ | 0.0013 | 0.0023 | 0.0060 |
| $w_9$ | 0.00042 | 0.00080 | 0.00260 |
| $w_{10}$ | 0.00012 | 0.00025 | 0.00100 |

**Table 2**: Weights $w_m$ for cutting $m$ Pomerons ($m$ colour exchanges).

where $\omega$ is the Fourier transform of the Pomeron propagator $G$,

$$\omega(s,b) = \frac{1}{i\pi} \int d^2k\, G(s,t)\, e^{-i\vec{k}\vec{b}}. \tag{4}$$

It should be noted that only two-dimensional integrations occur ($d^2b, d^2k$). The reason is that at high energies transferred momenta are purely transverse, and therefore longitudinal and transverse degrees of freedom decouple. Longitudinal integrations can be performed first, so the convolutions $G \otimes ... \otimes G$ mentioned earlier are simply two-dimensional integrations referring to transverse momenta. This is the reason why results can be presented in "impact parameter representation", namely in the form $\int d^2b...$.

Using the above result for $A(s,t)$, the elastic cross section is given as

$$\sigma_{\text{el}} = \int dt \frac{d\sigma_{\text{el}}}{dt} = \int d^2b\, |\gamma(s,b)|^2 \tag{5}$$

and, using the optical theorem, the total cross section can be written as

$$\sigma_{\text{tot}} = 8\pi\, \text{Im}\, A(s,0) = \int d^2b\, 2\text{Re}\gamma(s,b). \tag{6}$$

As a consequence, the inelastic cross section is given as

$$\sigma_{\text{inel}} = \sigma_{\text{tot}} - \sigma_{\text{el}} = \int d^2b\, \{1 - e^{-2\omega(s,b)}\}. \tag{7}$$

After having discussed elastic scattering, we would like to study inelastic amplitudes representing particle production. Unitarity relates elastic and inelastic amplitudes via where the sum extends over a complete set of final states. Since inelastic amplitudes cannot be calculated directly, one first treats the imaginary part of the elastic amplitude and then draws conclusions about inelastic scattering. For this purpose, the AGK technique is employed [9], which provides a method to express $\mathrm{Im}A(s,t)$ in terms of elementary inelastic scatterings, given as $\mathrm{Im}G(s,t)$, with $G$ being the Pomeron propagator. Using this technique and the optical theorem, the total cross section can be expanded as

$$\sigma_{tot} = \sum_{m=0}^{\infty} \sigma_m, \qquad (8)$$

where $\sigma_m$ is the cross section for $m$ elementary inelastic scatterings. The $\sigma_m$ can be calculated, they are weakly energy dependent, as shown in table 2, where we show $w_m = \sigma_m/\sigma_{\mathrm{inel}}$.

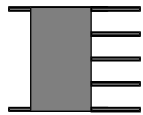

To be more specific about the "elementary inelastic processes", we need to know something about the nature of the Pomeron. We adopt Veneziano's picture of a Pomeron being a cylinder (of gluons and quark loops),

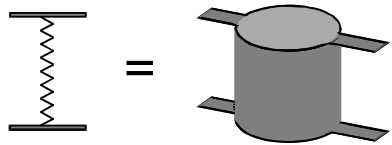

As a consequence, the imaginary part of the Pomeron amplitude is taken to be a squared "cut cylinder",

This is somehow the unitarity equation of a single Pomeron (compare with the full unitarity equation). Such a cut cylinder is still too complicated to be evaluated. It is, however, well known that planar QCD diagrams have surprisingly similar properties as compared to fragmenting strings [3], and correspondingly the cut cylinder is parametrized as two independently fragmenting relativistic strings. To indicate this

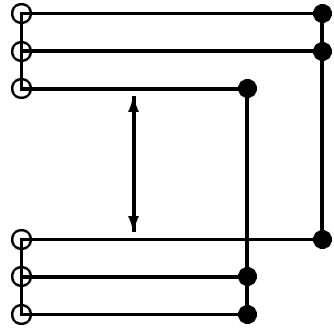

**Figure 3**: Quark line diagram.

and to keep track of the flavour flow, it turned out to be useful to to introduce quark line diagrams as shown in fig. 3 for a single colour exchange in the case of the VENUS model (Similar diagrams are also used in the DPM and QGSM). Quarks (and antiquarks) are represented by horizontal lines, and vertical connection of quark lines represent strings (including the incident hadrons). The quarks belonging to a string are highlighted by dots. This interaction process represents a colour exchange mechanism: before the interaction, the projectile and the target quarks are singlets, afterwards – as if colour would be exchanged between a projectile and a target

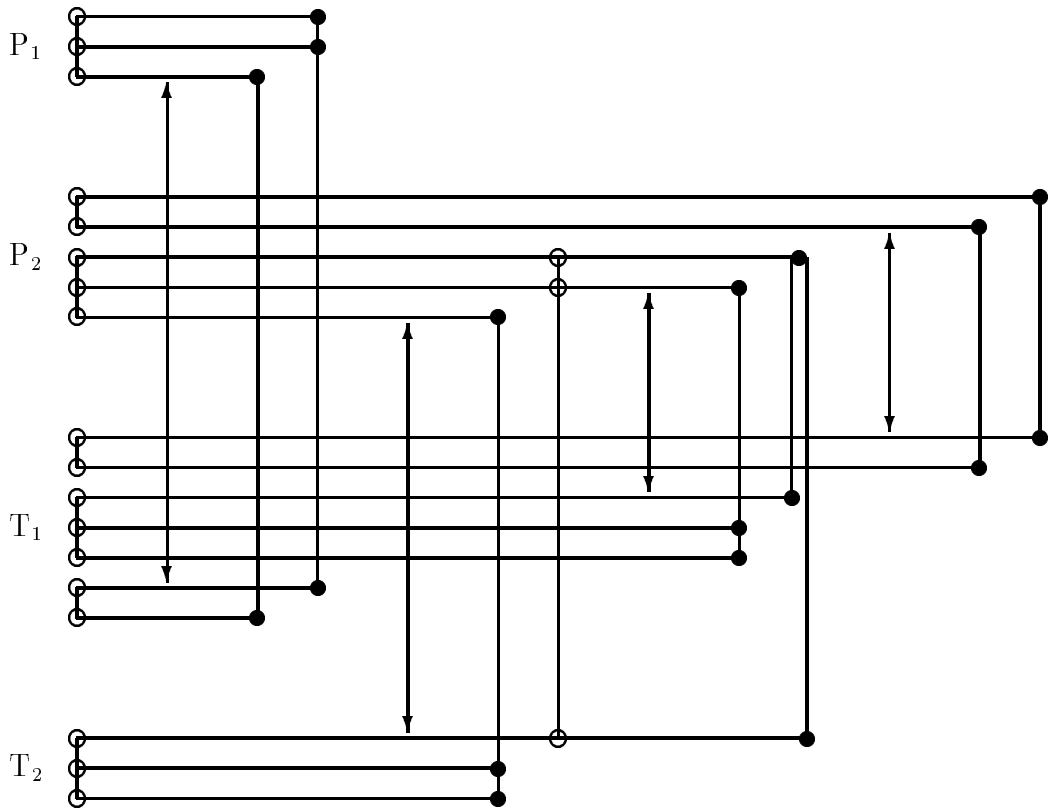

**Figure 4**: Quark line diagram for a nuclear interaction.

quark – a diquark forms a singlet together with a quark from the other nucleon. The colour exchange is indicated by an arrow in the figure. The generalization to multiple colour exchange is straightforward and obvious.

So far we have been discussing hadron-hadron scattering. The generalization to nucleus-nucleus is straightforward: starting from an expression for the elastic amplitude in terms of multiple Pomeron exchange, one obtains, by using the AGK technique, a multiple colour exchange picture for nuclear scattering. No new parameters need to be introduced, apart of those specifying the nuclear geometry. A simple quark line diagram for string formation in nuclear collisions is shown in fig. 4.

# 4 Limitations and New Directions

The model described so far is the so-called "independent string model", referring to the fact that the strings – even in nuclear collisions – fragment independent of each other. There are a couple of obvious limitations of the approach:

- **The low energy limit:** below some energy $\sqrt{s_{\min}}$ the string picture breaks down completely.

- **The high energy limit:** beyond some energy $\sqrt{s_{\max}}$ new components have to be added ($\rightarrow$ "soft & hard scattering")

- **The high density limit:** for high densities of produced hadrons, the concept of independent strings has to be modified ($\rightarrow$ "secondary interactions").

In the following, we discuss these limitations as well as new developments to go beyond.

## 4.1 Breakdown of the string model at low energies

Let us first introduce some time scales: the "hadronization time" $\tau_h$ is the time it takes before a hadron is formed, the "inter-collision time" $\tau_c$ is the time between successive nucleon-nucleon collisions, and the "traversal time" $\tau_R$ is the time it takes for a projectile nucleon to traverse a nucleus of radius $R$. To be specific and to simplify the discussion, we consider in the following a nucleon-nucleus collision, in the rest frame of the nucleus.

We may define an incident energy leading to

$$\tau_h < \tau_c$$

as "low energy". This relation implies that, after a nucleon-nucleon interaction, hadrons are produced essentially immediately before another target nucleon gets involved. This leads to the "cascade picture": the first nucleon-nucleon interaction produces a couple of hadrons, each of which – when hitting another target nucleon – will interact and produce hadrons itself, and so on. Such a scenario can be formulated completely in terms of hadron, without referring to the hadronic substructure.

The situation is completely different at high energy, defined via

$$\tau_h > \tau_R.$$

This means the following: after the first interaction, the leading "object" - the forward end of a string – traverses the whole nucleus, making interactions whenever a target nucleon comes in its way, before a leading hadron is formed outside the nucleus. This is actually the scenario employed by the string model approach, and therefore the relation $\tau_h > \tau_R$ may be used to estimate the energy $E_{min}$ below which the string model breaks down:

$$\tau_h = \gamma \tau_h^* = \frac{E}{m_N} \tau_h^* \approx E \frac{\text{fm}}{\text{GeV}} > \tau_R \approx 2R \approx 10 \text{fm}$$

which leads to

$$E > 10 \,\text{GeV} =: E_{\min},$$

or, in the cms,

$$s > s_{\min} \approx \sqrt{2 m_N E_{\min}} \approx 5 \,\text{GeV}$$

For energies less that that, we also have to deal with very low mass strings, adding another uncertainty. A third reason for not to use the string model below $s_{\min}$ is that the Gribov-Regge formalism, the basis of the string model, is based upon asymptotic considerations for $s/\text{GeV} \gg 1$. It appears to be most difficult if not impossible to fix all these shortcomings, so we consider $s_{\min} \approx 5 \,\text{GeV}$ an absolute lower limit, and do not attempt to go beyond.

## 4.2 Soft and hard scattering

From collider experiments at CERN and Fermilab we know that for energies ($\sqrt{s}$) beyond $10^2$ GeV a new feature shows up in nucleon-nucleon scattering: jets, referring to almost collinear bunches of hadrons. The origin of this phenomenon are elementary interactions of the type parton + parton → parton + parton, where each outgoing parton decays into a jet of hadrons. The inclusive cross section $\sigma_{\text{jet}}$ for jet production can be calculated by evaluating elementary QCD diagrams, and the result is – with some uncertainty due to an arbitrary $p_t$-cutoff – the following [10]: $\sigma_{\text{jet}}$ is zero up to $\sqrt{s} \approx 30 \,\text{GeV}$, increasing slowly up to $\sigma_{\text{jet}} \approx 5$ mb at $10^2$ GeV, and rising quickly with energy, exceeding 40 mb around $10^3$ GeV. The total cross section, on the other hand, rises slowly, from 40 mb at 30 GeV, via 45 mb at $10^2$ GeV, to around 70 mb at $10^3$ GeV. The value of $\sigma_{\text{jet}}$ relative to $\sigma_{\text{tot}}$ indicates the relative strength of the hard processes, leading to jets. For values of $s$ below 30-100 GeV, hard processes can be ignored. This provides an upper energy limit for the string model, representing pure soft interactions:

$$\sqrt{s_{\max}} \approx 30 - 100 \, \text{GeV}$$

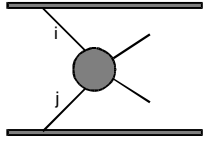 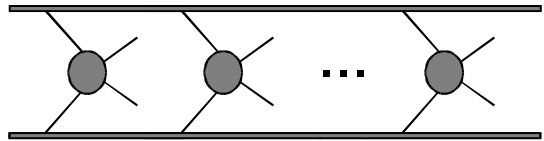

**Figure 5**: Hard scattering.                   **Figure 6**: Multiple hard scattering.

To treat higher energies, the string model needs to be extended to incorporate also hard processes.

Let us first briefly explain the standard perturbative QCD (PQCD) treatment of hard processes. Starting point is the following expression for the inclusive cross section for jet production [10, 11, 12, 13, 14, 15]:

$$\sigma_{\text{h}} = \sum_{ij} \int dx_1 \int dx_2 \int dt \frac{d\sigma_{ij}}{dt} f_i(x_1, Q^2) f_j(x_2, Q^2) \tag{9}$$

with the corresponding diagram shown in fig. 5. The indices $i$ and $j$ refer to partons of the nucleons (quarks/antiquarks of a certain flavour or gluons), $f_i$ and $f_j$ are the corresponding momentum distribution functions, with $x_1$ and $x_2$ being the longitudinal momentum fractions with respect to the incident momenta, and with $Q$ being the transferred momentum. The quantity $d\sigma_{ij}/dt$ represents the elementary cross section, calculated from summing QCD diagrams of lowest order.

It turned out that the inclusive cross section at high energies exceeds the inelastic cross section,

$$\sigma_{\text{h}} > \sigma_{\text{inel}}, \tag{10}$$

which reflects multiple scattering: The ratio

$$\bar{n} := \frac{\sigma_{\text{h}}}{\sigma_{\text{inel}}} \tag{11}$$

represents the average number of hard scatterings – which may be larger than one. In order to quantitatively reproduce not only $p_t$ spectra but also multiplicity distributions, an impact parameter model has been introduced [10, 11, 14]. Here one introduces an average number of collisions for a given impact parameter as

$$\bar{n}(b) := \sigma_{\text{h}} T(b) \tag{12}$$

with

$$T(b) := \int \rho(b - b') \rho(b'), \tag{13}$$

where
$$\rho(b) := \int dz\, \rho(b_x, b_y, z) \tag{14}$$

is the transverse matter density of a nucleon. The quantity $T(b)$ is a measure of the overlap of the two nucleons: $T(b)$ is largest for $b = 0$ (representing complete overlap), drops with increasing $b$, and approaches zero around $b = 2R_{\mathrm{nucleon}}$ (representing two touching nucleons).

Assuming independent multiple scattering, one writes the probability $\mathrm{prob}(j, b)$ of $j$ scatterings at given impact parameter as a Poissonian,

$$\mathrm{prob}(j, b) = \frac{\bar{n}(b)^j}{j!} e^{-\bar{n}(b)}, \tag{15}$$

with $\bar{n}(b)$ given in eq. (12) (see fig. 6). Summing over $j$ and integrating over $b$, one obtains the inelastic cross section,

$$\begin{aligned} \sigma_{\mathrm{inel}} &= \int d^2 b \sum_{j=1}^{\infty} \mathrm{prob}(j, b) \\ &= \int d^2 b \left\{ 1 - \exp[-\sigma_{\mathrm{h}} T(b)] \right\}. \end{aligned} \tag{16}$$

This is the well-known eikonal form.

We realize a large formal similarity between soft and hard scattering: in both cases multiple scattering leads to the "eikonal form" of total or inelastic cross section. For soft scattering, we have multiple cut Pomerons (multiple colour exchanges), for hard scattering, we have multiple jet production. Combining soft and hard amounts to considering a Pomeron (**P**) having a soft and a hard component,

$$\mathbf{P} = \mathbf{P}_{\mathrm{soft}} + \mathbf{P}_{\mathrm{hard}},$$

with a hard cut Pomeron representing a hard scattering and parton radiation. There are several attempts along theses lines [3, 7, 13], not being satisfactory though due to the necessity of an artificial low-$p_t$ cut-off. There is at present also a large uncertainty concerning predictions for multiplicities in nucleus-nucleus collision at RHIC or LHC energies.

## 4.3   Secondary Interactions

It is obvious that the independent string model has to break down, when the density of produced hadrons from string decay gets too large. This seems to happen already for moderately heavy nuclei $(S + S)$ at SPS energies.

Three approaches have been suggested to extend the independent string model correspondingly:

- the hadronic cascade;

- the string fusion mechanism;

- the formation of quark-matter clusters.

The <u>hadronic cascade</u> [7, 2] amounts to following the trajectories of hadrons from string decay and considering interactions according to tabulated cross sections, when two trajectories come sufficiently close. This method of purely hadronic binary interactions, however, obviously breaks down for densities such that more than two hadrons are likely to be closer than an interaction distance (which is probably the case already for S+S at SPS energy).

<u>String fusion</u> [2, 8, 16] considers secondary interactions already at an early stage, before particles are formed. In case of two strings being close together, they "fuse" to form a new string with increased charges at the endpoints. Two $\bar{3} - 3$ strings will fuse into a $\bar{6} - 6$ string (with weight 2/3) or a $3 - \bar{3}$ string (with weight 1/3), where the 3, $\bar{3}$, 6, and $\bar{6}$ refer to SU(3) representations, and the weights 2/3 and 1/3 result from the rules of multiplying SU(3) representations ($3 \times 3 = 6 + \bar{3}$). The break of a $6 - \bar{6}$ string requires at least a diquark-antidiquark production (which is rare for "normal strings") and will lead to a large baryon-antibaryon production rate. Details about string fusion may be found in [2, 8, 16].

<u>Formation of quark matter clusters</u> (QM–cluster approach) is the third approach to extend the independent string model [17], to be discussed in the following. Let us first introduce the basic ideas of the QM–cluster approach in a schematic way, the details and, in particular, the appropriate relativistic formulation will be given later. Consider a snapshot at some fixed time $\tau$. This time shold be large enough ($> 1$ fm/c), so that the "initial stage" interactions occuring in the nucleus-nucleus overlap zone are finished. At the given time $\tau$, we consider the locations (in $R^3$) of all particles produced in the initial stage. There are, by chance, regions with high density and such with low density. To be quantitative, we introduce a "critical density" $\rho_0$, and look for domains with $\rho > \rho_0$. The high density domains, the connected regions with $\rho > \rho_0$, are referred to as quark matter clusters (or droplets). Once they have been formed, these clusters are treated macroscopically, characterized by a distribution function for energy density $\varepsilon_i(\vec{x}, \tau)$, momentum $\vec{p}_i(\vec{x}, \tau)$, and flavour $f_i(\vec{x}, \tau)$, with $i$ referring to cluster $i$. There are no constituents explicitly kept track of, and there is no memory referring to the production process.

In order to determine high density regions at given time $\tau$, we proceed as follows: we assign a "critical volume" $V_0 = 1/\rho_0$ to all particles. Domains of high density correspond to overlapping volumes (see fig. 7). So the task here is to find connected objects with at least pairwise overlap of individual volumes (in fig. 7, we find four such objects). These objects are now considered as quark matter clusters (or droplets). It should be noted that the "critical volume" $V_0$ is not the usual volume $4\pi R_h^3/3$ of a hadron, it is rather defined by the requirement that if hadronic matter is compressed beyond $V_0$ per hadron, the individual hadrons cease to exist and quark matter is formed. So $V_0$ is less than $4\pi R_h^3/3$. Rather than the critical volume $V_0$ or the critical density $\rho_0$, we usually use the corresponding critical energy density $\varepsilon_0$.

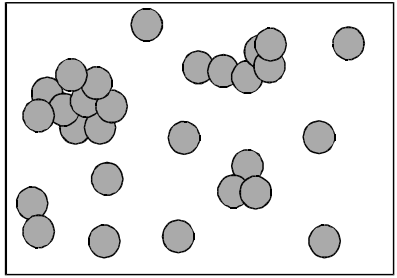
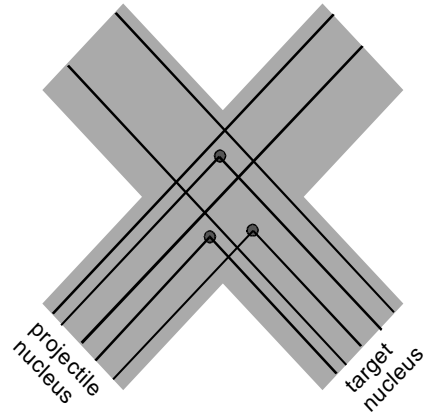

**Figure 7**: Overlapping volumes, representing high energy densities.

**Figure 8**: The collision zone (in the t–z plane) of an AA collision. Each dot represents an origin of a string evolution.

This critical energy density $\varepsilon_0$ (or equivalently $V_0$ or $\rho_0$) is the crucial parameter of our approach.

For the first stage, we use the independent string model, to be more precise the basic VENUS model without final state interactions [3]. In fig. 8, we show a typical event: we consider a projection to the $t$–$z$ plane, nucleon trajectories are represented by straight lines, interactions are represented by dots. Some nucleons interact (participants), some survive the interaction zone (spectators). Each dot, representing interaction, is a point of string formation, or, in other words, the origin of a string evolution. What happens after a string formation point? We use the standard procedure of classical relativistic string dynamics and decays [3]. In fig. 9 a typical example of the space-time evolution is shown. The upper rectangles represent produced particles (hadrons and resonances), the arrows indicate particle trajectories. Remarkable is the strict ordering of the directions, being a consequence of the covariant string breaking mechanism.

Being able to construct in a first stage, event by event, particle trajectories defined by their origins in space and the four-momenta, we can proceed to stage two, the analysis of energy densities and cluster formation at fixed time. The question is, what we mean by fixed time, which frame we are using. In this context, it is important to realize the empirically found correlation between the average rapidity $\bar{y}$ and the space-time rapidity $\zeta$,

$$\bar{y} = \zeta, \qquad (17)$$

with deviations only around $z \approx t$ due to the finite size of the nuclei. So there is an "inner region" with Bjorken-type behaviour $\bar{y} = \zeta$, where all the particle momenta point back to the origin, and an "outer region" with parallel velocity vectors. Correspondingly, the dotted line in fig. 10, a hyperbola in the inner region and the tangents at $z^P$ and $z^T$, represent equal proper time $\tau$ (on the average). In this way, obviously, also an average comoving frame is defined. The hypersurface defined by the hyperbola/tangents (dotted line) together with arbitrary $x_\perp$ is called $\tau$-hypersurface. We are now in a position to specify the frame for interactions: we investigate densities (or overlap) at constant $\tau$, which means on $\tau$-hypersurfaces.

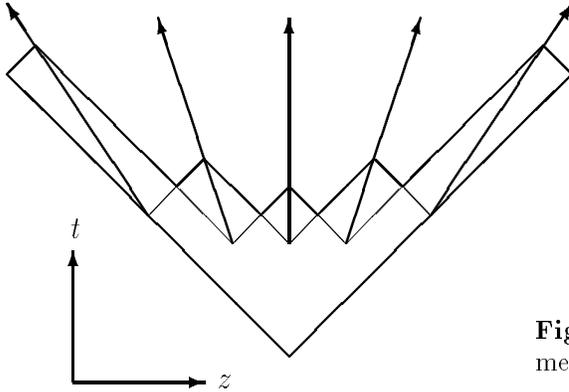

**Figure 9**: Trajectories of string fragments.

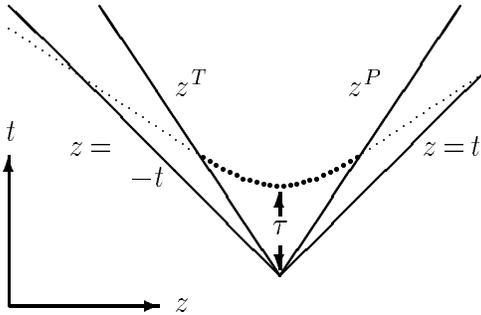

**Figure 10**: Space–time evolution of nucleus-nucleus scattering. The dotted line represents constant proper time (big dots: hyperbola, small dots: tangent).

Having specified the frame and correspondingly the time coordinate $\tau$, we have to introduce a useful longitudinal coordinate. We use the "proper length"

$$s := \int_0^z dz^*, \qquad (18)$$

with an integration at constant $\tau$, and with $dz^*$ being a longitudinal length in the average comoving frame defined by the $\tau$-hypersurface.

To specify the geometrical properties of particles or clusters, we use the variables $\tau$, $s$, and $\vec{r}_\perp = (r_x, r_y)$. At given $\tau$, a particle or cluster is considered as cylinder in $s, r_x, r_y$ space, with the axis along the $s$-axis. The object is characterized by a lower and upper value of $s$, $s_1$ and $s_2$, and a transverse radius $r_\perp$.

We are now in a shape to construct high density domains, to be considered as quark matter clusters. As discussed above, these domains are constructed, as in percolation models, by investigating geometrical overlaps of individual objects. There are two types of objects: initially, we have only hadrons and resonances, later also clusters contribute, which have been formed earlier. In any case, at a given time $\tau$, the objects are considered cylindric in $(s, r_x, r_y)$-space. Connected overlapping objects in this space define clusters. Such a cluster has in general a very irregular shape, which is "smoothened" in the sense that this complicated shape is replaced by a cylinder of the same volume, the same length $s_2 - s_1$, and the same center $(s_1 + s_2)/2$.

Starting at some initial time $\tau_0$ (presently 1 fm/c), we step through time as $\tau_{i+1} = \tau_i + \delta\tau$, constructing clusters at each time $\tau_i$. For the time evolution of clusters, we presently assume purely longitudinal expansion,

$$\Delta\zeta(\tau_{i+1}) = \Delta\zeta(\tau_i). \tag{19}$$

The last and most difficult topic to be discussed is the hadronization. The power of our percolation approach is that **any** hadronization scheme can be plugged into our approach and tested in a very detailed fashion. This is what we plan for the future. Currently, we present a very simple scenario, which can be implemented quite easily. Since our cluster expands, the energy density decreases, and drops at some stage below $\epsilon_0$. Per definition, we hadronize the cluster at this point instantaneously. Since the clusters turn out to have essentially longitudinal shape ($\Delta s \gg \Delta r_\perp$), we proceed as follows. The cluster is cut into many short pieces in $s$, all of them having the same mass $m$ with the requirement of $m$ being around some parameter $m_{\text{seg}}$. The small clusters then decay isotropically according to phase space [18]. So this is essentially the decay of many fireballs at different rapidities.

Crucial for the approach is the initialization, i.e., the volume $V_0$ assigned to the hadrons and resonances. The "critical" volume $V_0$, or equivalently the "critical" energy density $\epsilon_0$, is therefore a most important parameter. One believes nowadays, that the QCD phase transition is not first order, but rather a narrow transition within a range of about 10-20 MeV, corresponding to a $\epsilon$-range of maybe 0.1 to 2 GeV/fm$^3$. So the critical energy should be somewhere in this range, perhaps 1 GeV/fm$^3$. We calculate cluster-volume distributions for values of $\epsilon$ (CED) of 0.15, 0.50, and 1.0 GeV/fm$^3$, as shown in Fig. 11. The numbers are not normalized, we show the number of clusters per volume bin $\Delta V$, found in 1200 simulations; the bin sizes for the three distributions are, from top to bottom: $\Delta V$=10 fm$^3$, 20 fm$^3$, 60 fm$^3$. From the grey-scales, we can also read off how the clusters are distributed in energy density. We show only results for $\tau = 2$ fm/c, since volumes and energy densities turn out to scale in a trivial manner, as $V \sim \tau$ and $\varepsilon \sim \tau^{-1}$, so the distributions for different $\tau$ look similar up to a scale transformation.

We observe the following: for $\varepsilon_0 = 1$ GeV/fm$^3$, the distribution peaks at small values of $V$, dropping very fast with increasing volume $V$. Reducing the CED to 0.50 GeV/fm$^3$, the distribution gets wider, roughly by a factor of two. Although, as for $\varepsilon_0 = 1$ GeV/fm$^3$, very small clusters are favoured, the fluctuations are considerably larger. Reducing $\varepsilon_0$ further to 0.15 GeV/fm$^3$ ($\hat{=}$ nuclear matter density), we observe a drastic change in the distribution, a maximum at large values of $V$ emerged. This means, it is quite likely to observe just one big cluster (filling the whole volume).

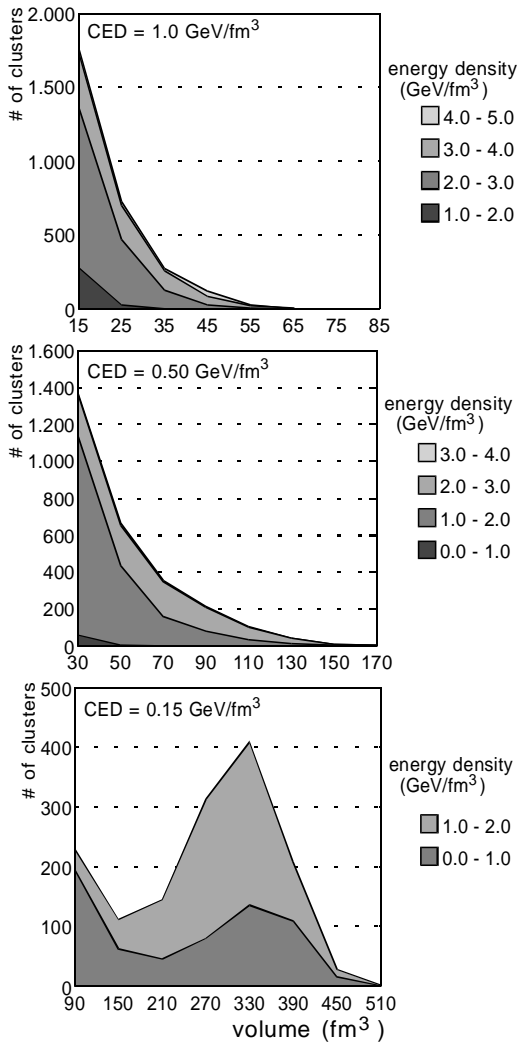

**Figure 11**: Distribution of cluster volumes for different values of the critical energy density (CED) $\epsilon_0$.

Considering $\varepsilon_0^* = 1$ GeV/fm$^3$ as the most realistic value, the upper plot of fig. 11 represents the realistic world (the lower plots are just mathematical exercise). Although here it is most likely to produce just small clusters (mainly hadrons and resonances), there is nevertheless a reasonable probability to form big clusters, with our statistics of 1200 simulations up to 65 fm$^3$. The question is, whether such "miniplasmas" can be "isolated" in event-by-event experiments.

## 5 Conclusions

We discussed the basic concepts of the string model, also to demonstrate that it is theoretically well justified. We outlined the limitations of the independent string model at low energies, at high energies, and at high particle densities. In this context, we reported on recent attempts to overcome these limitations by combining soft and hard scattering and by considering secondary interactions.